\documentclass[a4paper]{article}

\usepackage{INTERSPEECH2022}

\usepackage{graphicx}
\usepackage{kotex}
\usepackage{multicol}
\usepackage{array}
\usepackage{subcaption}
\usepackage{cite}
\title{Transfer Learning Framework for Low-Resource Text-to-Speech using\\ a Large-Scale Unlabeled Speech Corpus}
\name{Minchan Kim$^{1*}$, Myeonghun Jeong$^{1*}$, Byoung Jin Choi$^1$, Sunghwan Ahn$^1$, Joun Yeop Lee$^2$,\\ Nam Soo Kim$^1$}
\address{
  $^1$Department of Electrical and Computer Engineering and INMC, \\Seoul National University, Seoul, South Korea \\
  $^2$Samsung Research, Samsung Electronics, Republic of Korea
  }
\email{\{mckim, mhjeong, bjchoi, shahn\}@hi.snu.ac.kr, jounyeop.lee@samsung.com, nkim@snu.ac.kr}

\begin{document}

\maketitle
\def\thefootnote{*}\footnotetext{These authors contributed equally to this work.}
\def\thefootnote{\arabic{footnote}}

\begin{abstract}
Training a text-to-speech~(TTS) model requires a large scale text labeled speech corpus, which is troublesome to collect. In this paper, we propose a transfer learning framework for TTS that utilizes a large amount of unlabeled speech dataset for pre-training. By leveraging wav2vec2.0 representation, unlabeled speech can highly improve performance, especially in the lack of labeled speech. We also extend the proposed method to zero-shot multi-speaker TTS~(ZS-TTS). The experimental results verify the effectiveness of the proposed method in terms of naturalness, intelligibility, and speaker generalization. We highlight that the single speaker TTS model fine-tuned on the only 10 minutes of labeled dataset outperforms the other baselines, and the ZS-TTS model fine-tuned on the only 30 minutes of single speaker dataset can generate the voice of the arbitrary speaker, by pre-training on unlabeled multi-speaker speech corpus. 
\end{abstract}

\section{Introduction}
In recent years, text-to-speech~(TTS) models~\cite{shen2018natural,ren2020fastspeech,jeong2021diff,kim2020glow,kim2021conditional} have achieved dramatic improvement in various perspectives including naturalness, intelligibility, generation speed and controllability. However, one of the unresolved problems of TTS is that training these neural TTS models requires large amounts of text labeled speech corpora for high-quality speech generation. For example, LJSpeech~\cite{ito2017lj}, a public single speaker corpus for TTS, consists of more than 20 hours of delicately recorded speech with transcriptions. The difficulty and cost of collecting the labeled dataset can limit the application in various fields. In the zero-shot multi-speaker TTS~(ZS-TTS)~\cite{arik2018neural, cooper2020zero, min2021meta, casanova2021sc}, which synthesize voices of new speakers with only a few seconds of reference speech, it becomes more difficult to collect labeled dataset. The dataset should be composed of as many speakers as possible for better speaker generalization.\\
Meanwhile, transfer learning has been widely used in various fields, especially when the labeled dataset is insufficient. In transfer learning, a neural network model is firstly pre-trained for indirectly related objectives, and then the pre-trained model is used for parameter initialization of fine-tuning or feature extraction. For example, in computer vision~(CV), various image classification models~\cite{simonyan2014very, he2016deep, tan2019efficientnet, dosovitskiy2020image} trained on ImageNet~\cite{deng2009imagenet} are usually fine-tuned to classify the new classes with little examples. In natural language process~(NLP), self-supervised pre-training methods~\cite{devlin2018bert, liu2019roberta, yang2019xlnet} make it possible to exploit a large-scale unlabeled dataset, and significantly improve the generalization performance of downstream tasks such as machine translation and sentence classification. Following these works, several studies have attempted to apply transfer learning for speech data~\cite{oord2018representation, baevski2020wav2vec, hsu2021hubert}. For instance, wav2vec~2.0~\cite{baevski2020wav2vec} is used for low-resource speech recognition, even when there is no labeled dataset~\cite{baevski2021unsupervised}. In addition, several models utilize the self-supervised representations for speech analysis, disentanglement and voice conversion~\cite{polyak2021speech,choi2021neural,kreuk2021textless}.\\
Inspired by these works, we propose a novel transfer learning framework for TTS which employs large amounts of unlabeled speech for pre-training. The proposed method operates on VITS~\cite{kim2021conditional} architecture and leverages the wav2vec~2.0 representation to extract pseudo phoneme sequences, which are used for pre-training as a substitute of phoneme sequences. We also carefully designed the fine-tuning procedure in consideration of the role of each component and the training mechanism of normalizing flow.\\
The contributions of our work are as follows:
\begin{itemize}
\item To the best of our knowledge, it is the first approach that utilizes the unlabeled dataset for TTS based on the transfer learning framework.
\item For the single speaker TTS, we verify that the proposed pre-training significantly improved the generated speech quality, especially in the lack of labeled dataset. Only 10 minutes of labeled dataset is required for synthesizing high fidelity speech.
\item We extend the proposed framework to ZS-TTS. Pre-training on the large corpus, which consists of a large number of speakers, improves the speaker generalization on unseen speakers and even allows to generate the reference voices with only 30 minutes of a single speaker TTS dataset.
\end{itemize}

\begin{figure*}
 \centering
 \begin{subfigure}{0.67\columnwidth}
 \includegraphics[width=\columnwidth]{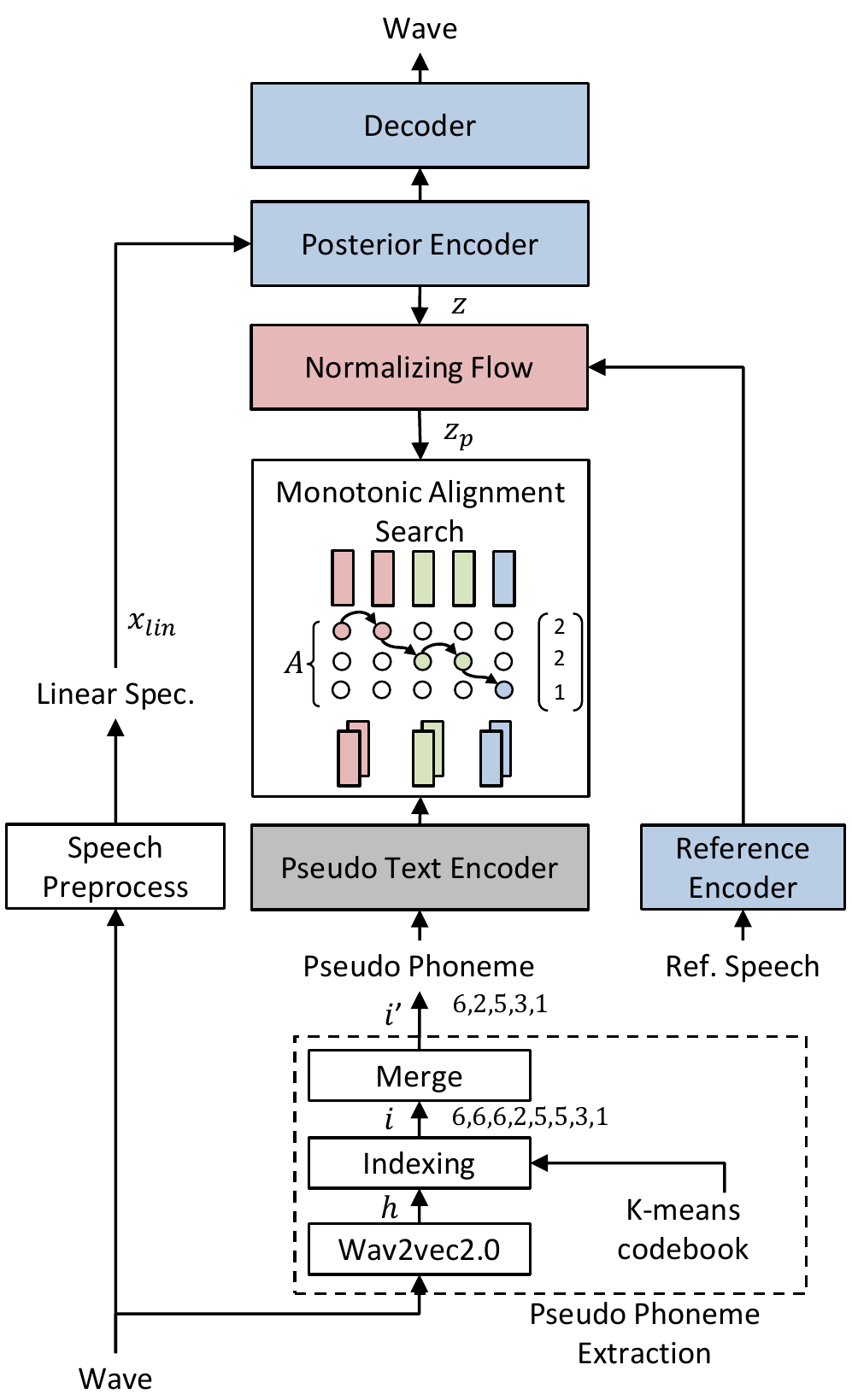}%
 \caption{Pre-training procedure}%
 \label{subfiga}%
 \end{subfigure}\hspace{0.2cm}
 \begin{subfigure}{0.67\columnwidth}
 \includegraphics[width=\columnwidth]{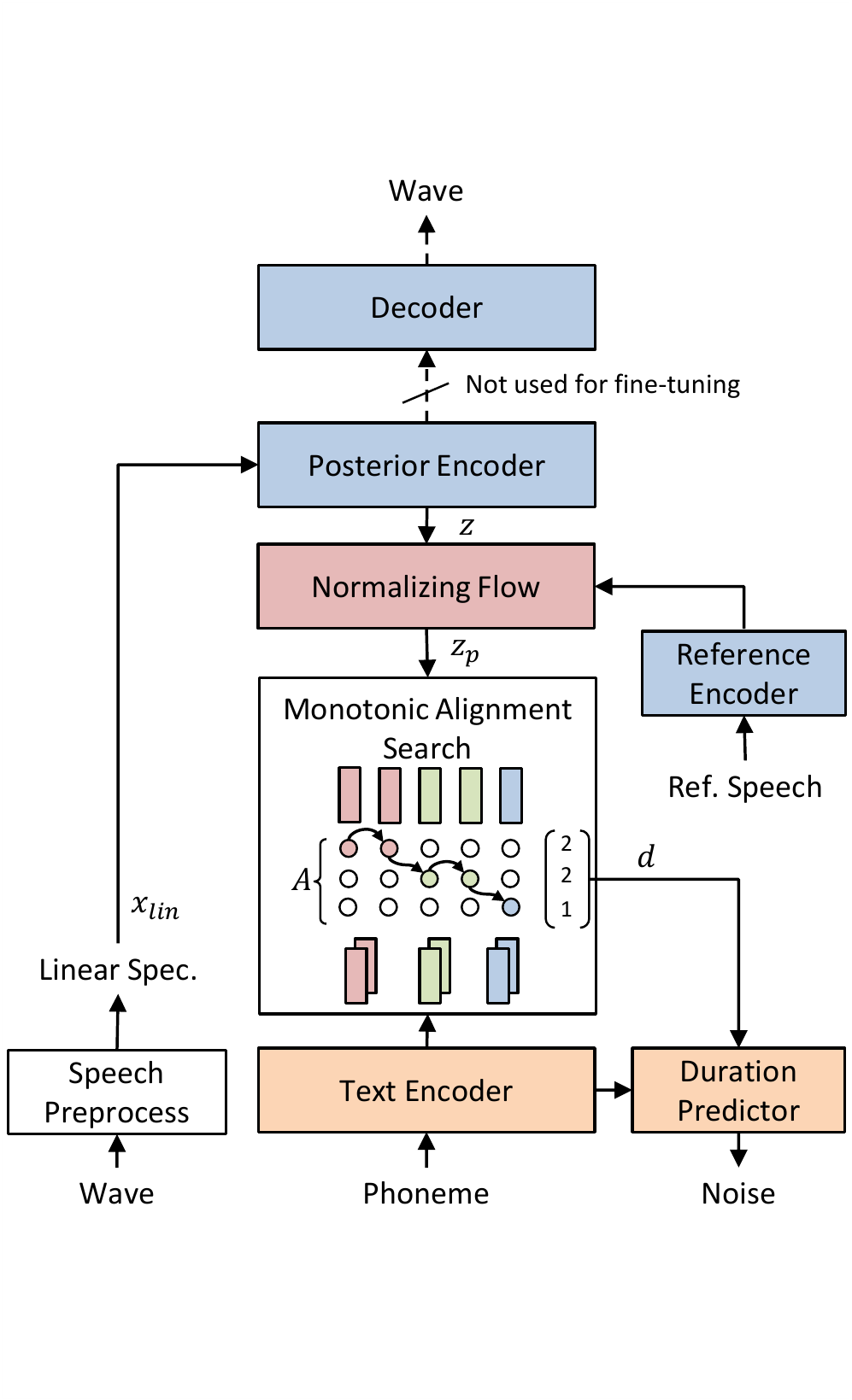}%
 \caption{Fine-tuning procedure}%
 \label{subfigb}%
 \end{subfigure}\hspace{0.2cm}
 \begin{subfigure}{0.67\columnwidth}
 \includegraphics[width=\columnwidth]{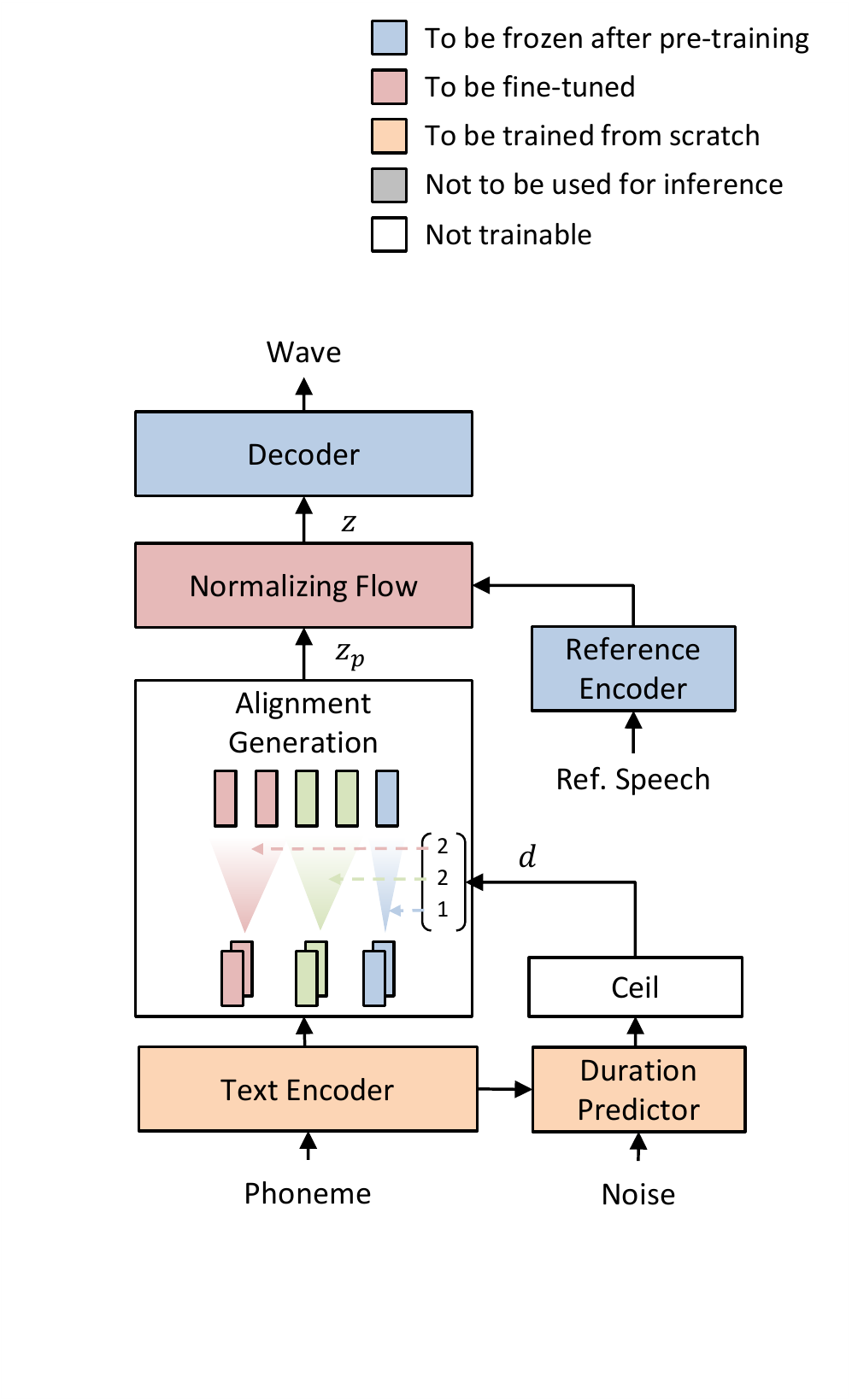}%
 \caption{Inference procedure}%
 \label{subfigc}%
 \end{subfigure}
 \caption{The entire framework of the proposed method. The discriminator part of the model is omitted for simplicity.} 
 \label{figure1}
\end{figure*}

\section{Proposed Method}
In this section, we describe the proposed transfer learning framework for TTS. As a base architecture, we use VITS~\cite{kim2021conditional} with several modifications. We first pre-train the VITS with a large untranscribed speech corpus and fine-tune the model with a small amount of text-labeled dataset. A new phonetic token named pseudo phoneme is used for pre-training as a substitute for phoneme. We explain pseudo phoneme in Section~\ref{pseudo phoneme}, and describe the detailed method in Section~\ref{transfer}. At last, we discuss the operation of the proposed framework in Section~\ref{discussion}. The overall proposed framework is depicted in Figure.~\ref{figure1}.

\subsection{Pseudo Phoneme}\label{pseudo phoneme}
Pseudo phoneme is a proposed token that contains phonetic information, and it is used as a substitute for a phoneme in pre-training. The pseudo phoneme should have similar characteristics with phoneme for effective pre-training and has to be acquired by speech-only corpus for our precondition of not using supervision. To satisfy these requirements, we leveraged the hidden representation of pre-trained wav2vec2.0 ~\cite{baevski2020wav2vec} which is trained in a self-supervised manner. According to ~\cite{baevski2021unsupervised}, k-means clustering on hidden representations of wav2vec2.0 operates well for assigning a phonetic token to each speech frame. The pseudo phoneme is extracted as follows. At first, we perform k-means clustering to identify $K$ clusters on the hidden representations of block 15 for the entire unlabeled speech corpus. We denote the hidden representation of block 15 as $h_{1},...,h_{T}$. Then, the cluster index $i_t \in \{1,...,K\}$ is obtained from each $h_t$, where we set $K=128$ for this work. As shown in Figure.\ref{subfiga}, the same consecutive indices are merged to reflect the characteristics of a real phoneme. We refer to these merged indices $i'_1,...,i'_{T'}$ as pseudo phoneme and use it for pre-training. In the above notation, $T$ and $T'$ denote lengths of speech frames and merged frames, respectively.

\subsection{Transfer Learning Framework for TTS}\label{transfer}
The proposed framework builds upon VITS architecture which has a conditional variational autoencoder (CVAE) structure. VITS consists of a posterior encoder, prior encoder, decoder and duration predictor. The posterior encoder extracts the posterior distribution $q_{\phi}(z|x_{lin})$ of the latent variable $z$ given the linear spectrogram $x_{lin}$, and the decoder generates output audio from the sampled $z$. The prior encoder consists of a normalizing flow and text encoder, and match the conditional prior distribution $p_{\theta}(z|c_{text}, A_{text})$ of $z$, given text $c_{text}$ and the alignment $A_{text}$. The alignment $A_{text}$ is calculated by monotonic alignment search~(MAS) algorithm~\cite{kim2020glow}, and the duration predictor learns the duration of each phoneme using $A_{text}$. Additionally, adversarial training helps to generate high fidelity sound. In the case of zero-shot multi-speaker TTS, we add a reference encoder with the same architecture of ~\cite{kim2021expressive} to extract speaker embedding. The speaker embedding is conditioned on the affine coupling layers of normalizing flow in the same way as ~\cite{kim2021conditional}, and it makes text-speaker dependent prior distribution of $z$.\\
\textbf{Pre-training}
For pre-training, we exploit the vanilla VITS training method except for two differences. At first, the pseudo phoneme and pseudo text encoder are used instead of phoneme and text encoder, respectively. The pseudo text encoder consists of 2 layers of 1D convolutions with a ReLU activation. Accordingly, the training objective is modified to maximize the conditional likelihood of speech given pseudo phoneme so that Kullback-Leibler divergence~(KLD) loss for prior matching is changed from Eq.~\ref{eq: kld1} to Eq.~\ref{eq: kld2}.
\begin{equation} \label{eq: kld1}
  L_{kl} = \log q_{\phi}(z|x_{lin}) - \log p_{\theta}(z|c_{text}, A_{text})
\end{equation}
\begin{equation} \label{eq: kld2}
  L_{kl'} = \log q_{\phi}(z|x_{lin}) - \log p_{\psi}(z|c_{i'}, A_{i'})
\end{equation}
In Eq.~\ref{eq: kld2}, ${\psi}$ represents parameters of the modified prior encoder: pseudo text encoder and normalizing flow, and $A_{i'}$ denotes the alignment between speech and pseudo phoneme. For ZS-TTS, the reference encoder is jointly trained with the other components.\\
\textbf{Fine-tuning}
The goal of fine-tuning is to adapt the pre-trained model to phoneme sequences. During fine-tuning, the pseudo phoneme and pseudo text encoder are replaced by phoneme and text encoder, and the KLD loss is also returned to Eq.~\ref{eq: kld1}. We carefully divide the model into three parts for efficient training: frozen, fine-tunned and scratch. Firstly, the decoder and posterior encoder are frozen during fine-tuning. As the role of the decoder and posterior encoder is to reconstruct high-quality raw audio, we assumed that these modules are trained enough with a large corpus and does not have to be fine-tuned with a small dataset. Freezing decoder induces several benefits. As we do not have to feed-forward to the frozen decoder, we can save lots of memory consumption and calculations for generating the high-resolution audio. In addition, there is no speech output during fine-tuning, so we can skip the adversarial training of VITS that makes the training process more complex. In the case of ZS-TTS, the reference encoder is also frozen during fine-tuning. We assume that fine-tuning reference encoders with small speakers can decrease the generalization performance for unseen speakers. Meanwhile, normalizing flow is fine-tuned to adjust the pseudo phoneme dependent prior to the phoneme dependent prior. It can mitigate the mismatch between the phoneme and pseudo phoneme. Lastly, the text encoder and duration predictor is trained from scratch. For better generalization, we do not condition speaker embedding to duration predictor. We briefly state that the only fine-tuned part is the normalizing flow, and only KLD loss and duration prediction loss are used for fine-tuning.

\subsection{Discussion on Framework Design} \label{discussion}

There is a notable difference between the existing transfer learning methods and the proposed framework. In transfer learning, a domain $\mathcal{D}$ is defined as $\mathcal{D}=\{\mathcal{X}, P(X)\}$, where $\mathcal{X}$ is a feature space and $P(X)$ is probability distribution of $X=\{x_{1},...,x_{n}\}\in {\mathcal{X}}$. Given a domain $\mathcal{D}$, a task is defined as $\mathcal{T}=\{\mathcal{Y}, f(x)\}$, where $\mathcal{Y}$ is a label space, and $f:\mathcal{X}\rightarrow\mathcal{Y}$ is an objective predictive function. The knowledge of source domain~$\mathcal{D}_S$ and source task~$\mathcal{T}_S$ can help training the target function~$f_T(x)$ of target task~$\mathcal{T}_T$ in target domain~$\mathcal{D}_T$.
However, in our framework, the source domain: pseudo phoneme indices and target domain: phoneme indices don't share any common feature spaces, meaning $\mathcal{X}_{S}\cap\mathcal{X}_{T}=\emptyset$. Instead, the $\mathcal{Y}_S$ and $\mathcal{Y}_T$ are highly related, since they are both speech. For this reason, feed-forward TTS models such as~\cite{ren2019fastspeech, ren2020fastspeech, lancucki2021fastpitch, vainer2020speedyspeech} are not suitable for the proposed method. When the domain changes, the text encoder should be randomly initialized because of feature space mismatch, and the decoder takes unlearned features from the text encoder. This makes it difficult to utilize the knowledge of the decoder acquired from pre-training and causes catastrophic forgetting~\cite{french1999catastrophic} during fine-tuning. To avoid this issue, we exploited the invertibility of normalizing flow. Normalizing flow has opposite directions for the training and inference procedure. In training procedure, the output of posterior encoder $z$ is converted to the latent variable of normalizing flow, $z_p$ $\sim\mathcal{N}(\mu_\theta(c_{text}, A_{text}), \sigma_\theta(c_{text}, A_{text}))$, where $\mu_\theta$ and $\sigma_\theta$ are calculated from text encoder. This property converts the domain shift problem to a target shift problem. Specifically, maximizing $\log p_{\theta}(z_p;\mu_\theta, \sigma_\theta)$ for the objective of Eq.~\ref{eq: kld1}\footnote{In Eq.\ref{eq: kld1}, $ p_{\theta}(z|c_{text}, A_{text})=\mathcal{N}(z_p;\mu_\theta, \sigma_\theta )|\mathrm{det}\frac{\partial z_p}{\partial z}|$.}, resembles minimizing weighted mean-squared-error~(MSE) between $z_p$ and $\mu_\theta(c_{text}, A_{text})$. This can be considered as target shift problem from $\mu_\psi(c_{i'}, A_{i'})$. As pseudo phoneme has rich phonetic information, the text encoder can quickly learn the well-defined latent space of $z_p$, and normalizing flow adapts for the gap between text and pseudo phoneme. 

\section{Experiments}
In this section, we verify the effectiveness of the proposed method in single-speaker TTS and zero-shot multi-speaker TTS. The detailed experimental setup and results of each case are described in Section~\ref{exp_single} and \ref{exp_multi}, respectively. Our synthesized audio samples are publicly available at https://jmhxxi.github.io/TransferTTS-demo/.

\subsection{Single Speaker TTS}\label{exp_single}
For single speaker TTS, we firstly pre-trained the single speaker VITS without transcription and fine-tuned on different sizes of the labeled dataset. This experiment demonstrates the effectiveness of the proposed method and its relation to the amount of the labeled dataset.\\
\textbf{Dataset}: We used LJSpeech~\cite{ito2017lj} dataset for single speaker TTS. The LJSpeech dataset consists of 13,100 short sentences, and the total length is about 24 hours with a sample rate of 22.05kHz. The dataset was randomly split into subsets of training set~(12500~sentences) and evaluation set~(500~sentences). All of the speech in the training set~(23~hours unlabeled) was used for pre-training, and different amounts of speech-text pairs in the training set were used for fine-tuning: LJ-10min~(10~minutes labeled), LJ-1h~(1~hour labeled), and LJ-10h~(10~hours labeled).\\
\textbf{Implementation Details}: We exploited the basic configuration of ~\cite{kim2021conditional} for the proposed model unless otherwise explained. We pre-trained the model for 500k iterations and fine-tuned it only for 30k iterations to generate high-quality speech. Both processes were trained with mini-batch size 64. For comparison, we used VITS-baseline, Glow-TTS~\cite{kim2020glow}, and FastSpeech2~\cite{ren2020fastspeech}. These baselines were trained on the fine-tuning subsets without pre-training. HiFi-GAN~\cite{kong2020hifi}, trained on the entire LJSpeech, was used as a vocoder for Glow-TTS and FastSpeech2.\\ 
\textbf{Evaluation Metrics}: For the subjective evaluation of audio fidelity, we conducted a mean opinion score~(MOS) test. 15 raters listened to the randomly selected samples and gave 5 scaled scores from 1 to 5 based on their naturalness. In addition, we estimated the character error rate~(CER) to evaluate the intelligibility of generated speech. We used a pre-trained speech recognition model from speechbrain toolkit~\cite{ravanelli2021speechbrain} for transcription.\\
\textbf{Experimental Results}: 
The results of single-speaker TTS are presented in Table~\ref{tab:single}. As shown in Table~\ref{tab:single}, the proposed method shows the best performance for both CER and MOS in all cases, and with the fewer labeled dataset, the larger the performance gap. This indicates that our proposed method using speech-only corpus significantly improve the performance, especially in the lack of the labeled dataset.

\begin{table}[th]
\setlength{\tabcolsep}{10pt}
\setlength{\arrayrulewidth}{0.2mm}
\caption{Comparison of CER and MOS with 95\% confidence intervals for single speaker TTS.}
\label{tab:single}
\centering
\begin{tabular}{l r r}
\toprule
\textbf{Method} & \multicolumn{1}{l}{\textbf{CER}}& \multicolumn{1}{l}{\textbf{\quad MOS}} \\
\midrule
Ground Truth        & 1.6 & 4.77$\pm$0.05   \\
\midrule
\textbf{LJSpeech 10min}\\
GlowTTS        & 31.6 & 1.82$\pm$0.11                                \\
FastSpeech2        & 15.3 & 1.89$\pm$0.11                               \\
VITS-baseline        & 8.0 & 2.20$\pm$0.10                                \\
Proposed        & \textbf{3.5} & \textbf{4.42$\pm$0.08}                             \\
\midrule
\textbf{LJSpeech 1hour}\\
GlowTTS        & 8.2 & 2.83$\pm$0.12                                \\
FastSpeech2        & 3.1 & 3.35$\pm$0.10                               \\
VITS-baseline        & 2.4 & 3.87$\pm$0.11                                \\
Proposed        & \textbf{2.3} & \textbf{4.58$\pm$0.07}                             \\
\midrule
\textbf{LJSpeech 10hour}\\
GlowTTS        & 2.1 & 4.37$\pm$0.09                                \\
FastSpeech2        & 2.6 & 3.90$\pm$0.09                               \\
VITS-baseline        & \textbf{1.9} & 4.60$\pm$0.07                                \\
Proposed       & \textbf{1.9} & \textbf{4.66$\pm$0.06}                             \\

\bottomrule
\end{tabular}
\end{table}

\subsection{Zero-Shot Multi-Speaker  TTS}\label{exp_multi}
Similar to the single speaker TTS, we pre-trained the zero-shot multi-speaker VITS model and fine-tuned it on different sizes of dataset. This experiment is conducted to verify that the proposed method improves the speaker generalization performance of ZS-TTS.\\
\textbf{Dataset:} For pre-training, we used train-clean-100 and train-clean-360 subsets of LibriTTS~\cite{zen2019libritts}. These subsets are about 245 hours in total, including 1151 speakers. In the case of fine-tuning, we used LJSpeech corpus and VCTK corpus~\cite{yamagishi2019cstr}. We made subsets of these corpus based on the total hours and number of speakers: LJ-30min~(30 minutes, 1 speaker), VCTK-1h~(1 hour, 4 speakers), VCTK-5h~(5 hours, 20 speakers) and VCTK-20h~(20 hours, 80 speakers). These subsets were used for fine-tuning. The evaluation set is composed of 28 speakers of VCTK not included in the fine-tuning sets so that all of the speakers in the evaluation set are unseen during training. The evaluation set has a total of 560 sentences and 20 sentences per speaker. All of the datasets were down-sampled to 22.05kHz for this experiment.\\
\textbf{Implementation Details:}
For ZS-TTS, the same configuration of the model was set with that of single speaker TTS except for the reference encoder. The proposed model was pre-trained for 500k iterations as in Section~\ref{exp_multi}, then fine-tuned for 10k iterations on LJ-30min and 100k iterations on the other subsets.
For comparison, we set the baselines with Meta-StyleSpeech~\cite{min2021meta}, SC-GlowTTS\footnote{We used a pre-trained speaker verification model~\cite{heo2020clova} for the reference encoder of SC-GlowTTS}~\cite{casanova2021sc}, and VITS-baseline. The VITS-baseline has the same architecture as the fine-tuned model but without pre-training, and these baselines were trained on the fine-tuning parts except LJ-30min. A pre-trained universal HiFi-GAN was used as a vocoder for Meta-StyleSpeech and SC-GlowTTS.\\
\textbf{Evaluation Metrics:} For speech quality and intelligibility, we used CER and MOS in the same way as Section~\ref{exp_single}. In addition, we evaluated the speaker similarity between reference speech and synthesized speech by similarity mean opinion score~(SMOS) and speaker embedding cosine similarity~(SECS). For SMOS, 15 raters evaluated whether generated speech reflects the speaker identity of reference speech well in the 5 scaled scores from 1 to 5. The SECS is the cosine distance of speaker embedding between generated speech and reference speech. To extract speaker embedding, we used a pre-trained speaker verification model~\cite{desplanques2020ecapa} from speechbrain toolkit. The SECS ranges from -1 to 1, and the higher score indicates the higher similarity.\\
\textbf{Experimental Results:} 
The results of ZS-TTS are presented in Table~\ref{tab:multi}. As shown in Table~\ref{tab:multi}, the overall performance gap between baselines and the proposed model increases as the datasets get smaller, and the performance of the proposed model is relatively less affected by the amount of fine-tuning dataset. 
Specifically, the proposed method achieves lower CER in all of the cases. Even though the MOS of the proposed model is lower than VITS-baseline in VCTK-5h and VCTK-20h, we expect that this gap can be mitigated by carefully matching the domain between pre-training and fine-tuning dataset. In the case of speaker similarity: SECS and SMOS, the proposed model significantly outperforms the baselines as well as VITS-baseline with any data sizes except for SC-GlowTTS of VCTK-20h. Since only SC-GlowTTS uses a pre-trained speaker verification model as a reference encoder, SC-GlowTTS might get higher SECS calculated by speaker verification model, while perceptual speaker similarity has a different tendency. One notable point is the results on LJ-30min. Although the proposed model is fine-tuned on a small single speaker dataset, it can generate voices of unseen speakers even with comparable performance. By this experiment, we show that the speaker similarity of ZS-TTS is heavily affected by the size of multi-speaker TTS corpus, and using large-scale unlabeled speech corpus can improve the speaker generalization performance, especially when the labeled dataset is limited.\\

\begin{table}[th]
\setlength{\tabcolsep}{2pt}
\setlength{\arrayrulewidth}{0.2mm}
\caption{Comparison of results for zero-shot multi-speaker TTS. MOS and SMOS is described with 95\% confidence intervals.}
\label{tab:multi}
\centering
\begin{tabular}{l r r r r}
\toprule
\textbf{Method} & \multicolumn{1}{l}{\textbf{CER}}& \multicolumn{1}{l}{\textbf{SECS}}& \multicolumn{1}{l}{\textbf{\quad MOS}}& \multicolumn{1}{l}{\textbf{\; SMOS}} \\ 
\midrule
Ground Truth        & 4.1 & 0.632 & 4.72$\pm$0.06 & 4.65$\pm$0.07  \\
\midrule
\textbf{LJ 30min, 1speaker}\\
Proposed      & 6.1 & 0.212 &3.99$\pm$0.10&3.47$\pm$0.09                                   \\
\midrule
\textbf{VCTK 1h, 4speakers}\\
SC-GlowTTS        & 29.5 & 0.151 &2.66$\pm$0.11&2.75$\pm$0.08                                \\
Meta-StyleSpeech        & 18.1 & 0.107 &2.12$\pm$0.11&1.98$\pm$0.09                                \\
VITS-baseline (+ ref.)       & 18.9 & 0.107 &3.36$\pm$0.11&2.84$\pm$0.08                                \\
Proposed        & \textbf{8.0} & \textbf{0.248} &\textbf{4.10$\pm$0.10}&\textbf{3.85$\pm$0.09}        \\
\midrule
\textbf{VCTK 5h, 20speakers}\\
SC-GlowTTS        & 14.0 & 0.248 &3.34$\pm$0.11&3.36$\pm$0.08                                \\
Meta-StyleSpeech        & 16.0 & 0.246 &3.52$\pm$0.12&3.69$\pm$0.09                                \\
VITS-baseline (+ ref.)       & 8.7 & 0.179 &\textbf{4.40$\pm$0.08}&3.60$\pm$0.09                               \\
Proposed        & \textbf{8.5} & \textbf{0.298} &4.21$\pm$0.10&\textbf{3.92$\pm$0.09}                   \\
\midrule
\textbf{VCTK 20h, 80speakers}\\
SC-GlowTTS        & 6.2 & \textbf{0.327} &3.94$\pm$0.09&3.79$\pm$0.09                                \\
Meta-StyleSpeech        & 14.4 & 0.302 &3.42$\pm$0.11&3.85$\pm$0.08                                \\
VITS-baseline (+ ref.)       & \textbf{5.7} & 0.267 &\textbf{4.44$\pm$0.08}&4.00$\pm$0.09               \\
Proposed        & \textbf{5.7} & 0.325 &4.32$\pm$0.09&\textbf{4.15$\pm$0.08}                                \\
\bottomrule
\end{tabular}
\end{table}

\section{Conclusions}
In this work, we introduced a transfer learning framework for TTS that uses a large amount of unlabeled speech dataset for pre-training. This framework is based on VITS architecture and the pseudo phoneme extracted from wav2vec2.0 representation. Due to the similar nature of phoneme and pseudo phoneme, we can pre-train the TTS model on a speech-only corpus, which improves the overall performance, especially in the lack of labeled speech. Our experimental results show that the proposed method outperforms other baselines over all of the fine-tuning subsets in single-speaker TTS and makes higher speaker similarity for zero-shot multi-speaker TTS. This performance improvement was greater with the smaller labeled dataset. With these advantages, we expect that the proposed method becomes a good starting point for low-resource TTS.

\section{Acknowledgements}
This work is/was supported by Samsung Research, Samsung Electronics
Co.,Ltd.

\bibliographystyle{IEEEtran}

\bibliography{mybib}
\end{document}